\documentstyle[12pt]{article}
\begin{document}
\title{Chiral anomaly and unambiguous rational terms}
\author{Jifeng Yang$^*$  and Guang-jiong Ni\\
Department of Physics, Fudan University, Shanghai, 200433, P. R. China}
\footnotetext{$^*$ E-mail:jfyang@fudan.edu.cn}
\maketitle
\begin{abstract}
Through more detailed calculations on QED$_{1+1}$ and QED$_{3+1}$, we
attribute the regularization independent and hence definite origin
of chiral anomaly in perturbation theory to an unambiguous term 
which is
a rational function in momentum space. Some relevant remarks are
presented.
\end{abstract}
\newpage
In spite of the great success of the geometrical method \cite{Ja} 
in the study of chiral anomaly \cite{ABJ}, we feel it still 
worthwhile to work out
explicitly the definite (or regularization independent) origin of
anomaly within the framework of quantum field theory (QFT). 
Presently, all the analysis, both perturbative ones and 
nonperturbative ones, are forced to employ specific 
regularizations, while the appearance of anomaly should be 
regularization independent within the QFT framework. 
There must be definite or more physical reasons behind the 
ambiguities caused by ultraviolet divergences. To our knowledge, 
satisfactory explanation is not available by now. One may find 
that the old lore that regularization originates anomaly still 
dominates the current literrature. Recently, two new 
regularization methods \cite{DF} are used to study the 
problem, but the definite origin of anomaly is still unclear 
there. At least, it is not explicitly seen there.

Our study here is motivated by a desire to find out regularization
independent or definite reason for the appearance of anomaly. It is
in this way can we convince ourselves whether the anomaly is a true
quantum mechanical effect (definite) or merely a regularization 
effect. We will use a simple approach. In fact, given a divergent 
Feynman integral, we may treat it in the following way ({\em 
before affording a specific regularization}) \cite {YN,JF}: First,
differentiate it with respect to the external momenta to arrive at 
a convergent (well-defined) one and perform this new integral. 
Then, proceed to integrate it with respect to the external momenta 
for going back to the original one. At last, one obtains an 
indefinite integral which consists of a well-defined function of 
external momenta plus a polynomial with unknown constants as 
coefficients. The divergences or ambiguities just reside in these 
constants. A specific regularization is just to afford definitions 
for these constants. {\em This leads to a regularization 
independent treatment of Feynman integrals, since we have
not done anything special to the integrals and no particular 
parameter is introduced} (or in a more conservative attitude, this 
procedure may be seen as providing a universal "regularization".
For more consistent description of this approach, please refer to 
\cite {JF}.)

With the above preparations, we proceed to study the relatively 
simpler case of QED$_{1+1}$ first. The amplitude is
\begin{eqnarray}
{\Gamma}^5_{\mu\nu} (p,m)
&=& - e \int\displaystyle\frac{d^2 Q}{(2\pi)^2}
    Tr\left ( {\gamma}_{\mu} {\gamma}_5
    \displaystyle\frac{1}{\not \/\!\! Q +
    \not \/\!p-m}{\gamma}_{\nu}
    \displaystyle\frac{1}{\not \/\!\!Q-m}\right ) \nonumber\\
&=& 2ie\left \{ {\epsilon}_{\alpha\mu} (I_{\nu}p^{\alpha} + 
   I^{\alpha} p_{\nu} +2I^{\alpha}{}_{\nu})+
    {\epsilon}_{\nu\mu}(I_{sq}+p\cdot I
    -m^2 I_0 )\right \}, \nonumber\\
{\Gamma}^5_{\nu} (p,m)
&=& - e \int\displaystyle\frac{d^2 Q}{(2\pi)^2}Tr\left ({\gamma}_5
    \displaystyle\frac{1}{\not \/\!\! Q+\not \/\!p-m}{\gamma}_{\nu}
    \displaystyle\frac{1}{\not\/\!\!Q-m}\right )
        =- 2iem{\epsilon}_{\nu\alpha} p^{\alpha}I_0 ,
\end{eqnarray}
where
\begin{eqnarray}
I_0 (p,m)
&=& \int\displaystyle\frac{d^2Q}{(2\pi)^2}
    \displaystyle\frac{1}{[Q^2-m^2][(Q+p)^2-m^2]}=
    -\displaystyle\frac{i}{2\pi\theta p^2}\ln
    \displaystyle\frac{\theta +1}{\theta -1},\nonumber\\
I_{\alpha} (p,m)
&=& \int \displaystyle\frac{d^2Q}{(2\pi)^2}
    \displaystyle\frac{Q_{\alpha}}{[Q^2-m^2][(Q+p)^2-m^2]}=
    \displaystyle\frac{ip_{\alpha}}{4\pi\theta p^2}\ln
    \displaystyle\frac{\theta+1}{\theta-1},\nonumber\\
I_{sq} (p,m)
&=& \int\displaystyle\frac{d^2Q}{(2\pi)^2}
    \displaystyle\frac{Q^2}{[Q^2-m^2][(Q+p)^2-m^2]}=
    \displaystyle\frac{i({\theta}^2-1)}{8\pi\theta}\ln
    \displaystyle\frac{\theta+1}{\theta-1}+c_1,\nonumber\\
I_{\mu\nu} (p,m)
&=& \int\displaystyle\frac{d^2Q}{(2\pi)^2}
    \displaystyle\frac{Q_{\mu}Q_{\nu}}{[Q^2-m^2]
     [(Q+p)^2-m^2]} \nonumber\\
&=& \displaystyle\frac{ip_{\mu}p_{\nu}}{4\pi p^2}
    \left ( 1-\displaystyle\frac{{\theta}^2+1}{2\theta}\ln
    \displaystyle\frac{\theta+1}{\theta-1}\right )+
    \displaystyle\frac{ig_{\mu\nu}\theta}{8\pi}\ln
    \displaystyle\frac{\theta+1}{\theta-1}+
    g_{\mu\nu}c_2,\nonumber\\
\theta
&=& \sqrt{1-4m^2/p^2}.
\end{eqnarray}
The integrals in Eq.(2) have been performed in the way mentioned 
above, where two unknown constants $c_1,c_2$ represent the 
ambiguities or divergences (or ill-definedness) of 
$I_{sq},I_{\mu\nu}$ respectively. Substituting Eq.(2) into Eq.(1) 
we have
\begin{eqnarray}
{\Gamma}^5_{\mu\nu} (p,m)
&=& -e(g_{\nu}{}^{\alpha}-
    p_{\nu}p^{\alpha}/p^2){\epsilon}_{\alpha\mu}
   \displaystyle\frac{\theta^2-1}{2\pi\theta}\ln
    \displaystyle\frac{\theta+1}{\theta-1}-e \epsilon_{\alpha\mu}
    p_{\nu}p^{\alpha}/(\pi p^2)+
     2ie{\epsilon}_{\mu\nu}(c_1-2c_2),\nonumber\\
{\Gamma}^5_{\nu} (p,m)
&=& e{\epsilon}_{\nu\alpha}p^{\alpha}
    \displaystyle\frac{{\theta}^2-1}{4\pi m\theta}\ln
    \displaystyle\frac{\theta+1}{\theta-1}.
\end{eqnarray}
Note that the expressions in Eq.(3) are regularization independent. 
From Eq.(3)we see
\begin{eqnarray}
p^{\mu}{\Gamma}^5_{\mu\nu} (p,m)
&=&- 2m {\Gamma}^5_{\nu} (p,m) -
    c_{\mu\mu}{\epsilon}_{\nu\alpha}p^{\alpha},\nonumber\\
p^{\nu}{\Gamma}^5_{\mu\nu} (p,m)
&=& -c_{\nu\nu} {\epsilon}_{\alpha\mu}p^{\alpha},
\end{eqnarray}
where
\begin{eqnarray}
& &c_{\mu\mu}= 2ie(c_1-2c_2), \ \  c_{\nu\nu}=
      2ie(c_1-2c_2)+e/{\pi},\\
& &c_{\nu\nu}-c_{\mu\mu}= e/{\pi}.
\end{eqnarray}
We stress that Eq.(6) is valid in any regularization. It is this
equation which tells us that one can not set $c_{\mu\mu}$ and
$c_{\nu\nu}$ to zero at the same time.The Eq.(6) is resulted by 
the term  $ip_{\mu}p_{\nu}/(4\pi p^2)$ in the ambiguous 
$I_{\mu\nu}$ (see Eq.(2)), or the 
$-e{\epsilon}_{\alpha\mu}p^{\alpha}p_{\nu}/(\pi p^2)$ term in
${\Gamma}^5_{\mu\nu}$ (see Eq.(3)). It is a nonlocal rational 
term. {\em Since the validity of eq.(6) is independent of 
regularization, the anomaly is inevitable.} Of course the 
${\epsilon}_{\mu\nu}$ tensor is indispensable in this problem, but 
it is only part of it, for it exists in every part of the
relevant amplitude, and some contribute to anomaly while some do
not.

Now we look at the $1+3$-dimensional (massless) case where
\begin{eqnarray}
{\Gamma}^5_{\rho\nu\mu} (q,p)
&=&{\Gamma}^{5(1)}_{\rho\nu\mu}(q,p)+
    {\Gamma}^{5(2)}_{\nu\rho\mu}(p,q)=
   2{\Gamma}^{5(1)}_{\rho\nu\mu}(q,p) \nonumber\\
&=& -2e^2\int\displaystyle\frac{d^4Q}{(2\pi)^4}Tr
    \left ( {\gamma}_{\rho}
    \displaystyle\frac{1}{\not \/\!\!Q} {\gamma}_{\nu}
    \displaystyle\frac{1}{\not \/\!\!Q-\not \/\!p} {\gamma}_{\mu}
    {\gamma}_5 \displaystyle\frac{1}{\not \/\!\!Q+
     \not \/\!q} \right )  \nonumber\\
&=&-8ie^2 \left \{ {\epsilon}_{\alpha\beta\nu\mu} 
    (2I_{\rho}^{\alpha} p^{\beta}+I_{\rho}q^{\alpha}p^{\beta}+
    p_{\rho}I^{\alpha}q^{\beta}+
    q_{\rho}I^{\alpha}p^{\beta}) \right . \nonumber\\
& &\ \ \ \ \  +{\epsilon}_{\alpha\beta\rho\mu} (2I_{\nu}^{\alpha}
    q^{\beta}+I_{\nu}q^{\alpha}p^{\beta}- 
    q_{\nu}I^{\alpha}p^{\beta}-
    p_{\nu}I^{\alpha}q^{\beta}) \nonumber\\
& &\ \ \ \ \  +{\epsilon}_{\alpha\beta\rho\nu} (-I_{\mu}q^{\alpha}
    p^{\beta}+q_{\mu}I^{\alpha}p^{\beta}-
    p_{\mu}I^{\alpha}q^{\beta})
    \nonumber\\
& &\ \ \ \ \  \left .+{\epsilon}_{\alpha\rho\nu\mu}  
    (I_{sq}p^{\alpha}+
    q\cdot Ip^{\alpha} -I_{sq}q^{\alpha}+ p\cdot Iq^{\alpha}
    -I_{sq}^{\alpha}-q\cdot pI^{\alpha})\right \}
\end{eqnarray}
where
\begin{eqnarray}
I_{\alpha} (q,p)
&=& \int \displaystyle\frac{d^4Q}{(2\pi)^4}
    \displaystyle\frac{Q_{\alpha}}{A_1 A_2 A_3},\ \ \ 
    I_{sq} (q,p)=
    \int\displaystyle\frac{d^4Q}{(2\pi)^4}
    \displaystyle\frac{Q^2}{A_1 A_2 A_3},\nonumber\\
I_{\mu\nu} (q,p)
&=& \int \displaystyle\frac{d^4Q}{(2\pi)^4}
   \displaystyle\frac{Q_{\mu}Q_{\nu}}{A_1 A_2 A_3},\ \ \ 
   I_{sq,\alpha}
  (q,p) = \int\displaystyle\frac{d^4Q}{(2\pi)^4}
    \displaystyle\frac{Q^2 Q_{\alpha}}{A_1 A_2 A_3}, \nonumber\\
A_1
&=& Q^2, \ \ \ A_2=(Q-p)^2, \ \ \ A_3=(Q+q)^2.
\end{eqnarray}
The integrals in Eq.(8) are also performed in the way described 
above. $I_{\alpha}$ is convergent, whereas the other three 
integrals are ill-defined, and three unknown $C_1$, $C_2$ and 
$C_3$ appear in the ambiguous $I_{sq}$, $I_{\mu\nu}$ and 
$I_{sq,\alpha}$ respectively. Since the expressions are lengthy, 
we do not list them here. The main point lies in the fact that 
there is also a rational term in $I_{\mu\nu}$
(a logarithmically divergent integral),
i.e. $-ik_{\mu}k_{\nu}/(2(4\pi)^2 k^2)$, with $k=p+q$, which will 
be responsible for the appearance of anomaly. Then,
from Eq.(7) and Eq.(8) we have
\begin{eqnarray}
{\Gamma}^5_{\rho\nu\mu} (q,p)
&=&-\displaystyle\frac{e^2}{2\pi^2}\epsilon_{\alpha\beta\nu\mu}
   q^{\alpha}p^{\beta}\left \{
   \displaystyle\frac{k_{\rho}}{k^2} +N_{0,0}\left [
   \displaystyle\frac{(p^2-q^2)^2 k_{\rho} +
   (p^2-q^2)k^2 q_{\rho}}{k^6}
   \right ] \right .\nonumber\\
& & -\displaystyle\frac{\ln(p^2/q^2)}{2k^4}
   \left [ -2p\cdot kq_{\rho}+
    (q^2-p^2)p_{\rho}\right ]  \nonumber\\
& & -\displaystyle\frac{N_{0,1}}{k^6}\left [
    2(p^4-p^2q^2+p\cdot kk^2-2p\cdot kq^2)q_{\rho}+
   (q^2 k^2+p^2 k^2-4q\cdot kp^2-4p\cdot kq^2)p_{\rho}\right ]
   \nonumber\\
& & + \left . \displaystyle\frac{2N_{0,2}}{k^6}\left [
   (\gamma^2+(p\cdot k)^2)q_{\rho}+
    (\gamma^2-p\cdot k q\cdot k)p_{\rho}
   \right ] \right \} \nonumber\\
& &-\displaystyle\frac{e^2}{2\pi^2}\epsilon_{\alpha\beta\rho\mu}
    p^{\alpha}q^{\beta}\left \{
   \displaystyle\frac{k_{\nu}}{k^2}+N_{0,0}\left [
   \displaystyle\frac{(p^2-q^2)^2 k_{\nu} +
    (q^2-p^2)k^2 p_{\nu}}{k^6}
   \right ] \right .\nonumber\\
& &-\displaystyle\frac{\ln(q^2/p^2)}{2k^4} \left[
   -2q\cdot kp_{\nu}-(q^2-p^2)q_{\nu}\right ]  \nonumber\\
& & -\displaystyle\frac{N_{0,1}}{k^6}\left [
    2(q^4-p^2 q^2+q\cdot kk^2-2q\cdot kp^2)p_{\nu}+
    (q^2 k^2+p^2 k^2-4p\cdot kq^2-4q\cdot kp^2)q_{\nu}\right ]
    \nonumber\\
& & +\left .\displaystyle\frac{2N_{0,2}}{k^6}\left [
    (\gamma^2+(q\cdot k)^2)p_{\nu}+
    (\gamma^2-p\cdot kq\cdot k)q_{\nu}
    \right ] \right \}\nonumber\\
& &-\displaystyle\frac{e^2}{4\pi^2}\epsilon_{\alpha\rho\nu\mu}
   (p-q)^{\alpha}\left \{ 4C_2+2C_3-2C_1
   -\displaystyle\frac{2p\cdot k\ln q^2/\sigma+2q\cdot k
    \ln p^2/\sigma}{k^2} \right .\nonumber\\
& & \left .+\ln k^2/\sigma+\displaystyle\frac{(p^2-q^2)^2}{k^6}
   N_{0,0}-\displaystyle\frac{4\gamma^2}{k^4}N_{0,2}
   +\displaystyle\frac{4p\cdot qk^2-8\gamma^2}{k^4}N_{0,1} 
    \right \}
    \nonumber\\
& &+\displaystyle\frac{e^2}{4\pi^2}\epsilon_{\alpha\rho\nu\mu}
   \left \{ \displaystyle\frac{k^2\ln q^2/p^2 +
   (q^2-p^2)N_{0,0}}{k^4}
    (q^2 p^{\alpha}+p^2 q^{\alpha}) \right .\nonumber\\
& &\left .+\displaystyle\frac{2N_{0,1}}{k^4}
    (p^2 q\cdot kq^{\alpha}-q^2 p\cdot kp^{\alpha})\right \}.
\end{eqnarray}
Here,
\begin{eqnarray}
N_{0,n}
&=&\int_0^1 dx\int_0^x dy \displaystyle\frac{x^n}{y^2+Cy+D},\ \
   n=0,\ 1,\ 2, \nonumber\\
C
&=& \displaystyle\frac{p^2-q^2-2x p\cdot k}{k^2},\ \ \
   D=\displaystyle\frac{p^2}{k^2}(x^2-x),\nonumber\\
\gamma^2
&=& (p\cdot q)^2-p^2q^2,\ \ \ \ k=p+q,
\end{eqnarray}
and $\sigma$ is a parameter of mass dimension two, which is 
irrelevant to the anomaly problem. We emphasize here that the 
rational terms 
$-\frac{e^2}{2\pi^2}\epsilon_{\alpha\beta\nu\mu}p^{\alpha}q^{\beta}
k_{\rho}/k^2$ and $-\frac{e^2}{2\pi^2}\epsilon_{\alpha\beta\rho\mu}
q^{\alpha}p^{\beta}k_{\nu}/k^2$ come from the rational term 
$-ik_{\mu}k_{\nu}/(32\pi^2 k^2)$ in $I_{\mu\nu} (q,p)$. From 
Eq.(9) we have the following equations
\begin{eqnarray}
q^{\rho}\Gamma^5_{\rho\nu\mu}
&=& \frac{e^2}{4\pi^2}\epsilon_{\alpha\beta\nu\mu}
    p^{\alpha}q^{\beta}R_1 (q,p,C_1,C_2,C_3),
   \nonumber\\
p^{\nu}\Gamma^5_{\rho\nu\mu}
&=& \frac{e^2}{4\pi^2}\epsilon_{\alpha\beta\rho\mu}
    q^{\alpha}p^{\beta}R_2 (q,p,C_1,C_2,C_3),\nonumber\\
k^{\mu}\Gamma^5_{\rho\nu\mu}
&=& \frac{e^2}{4\pi^2}\epsilon_{\alpha\beta\rho\nu}
    q^{\alpha}p^{\beta}R_3 (q,p,C_1,C_2,C_3).
\end{eqnarray}
$R_1$, $R_2$ and $R_3$ are expressions of $p$, $q$ and the three 
unknown constants $C_1$, $C_2$ and $C_3$. They satisfy the 
following equation
\begin{equation}
R_1+R_2+R_3=2.
\end{equation}
This is what we are seeking for. It is valid independent of
regularization, that is, we have obtained it without referring to 
any specific regularization scheme \cite {YN}. ($R_1$, $R_2$ and 
$R_3$ must be constants after one has carried out all the 
calculations in Eq.(9) and Eq.(10), otherwise nonlocal anomalies 
would appear.){\em  Eq.(12) implies that there is no way to make 
the right hand sides of equations in Eq.(11) vanish at the same 
time, then we encounter the anomaly}. If the rational term were 
missing in $I_{\mu\nu}$ one would get
\begin{equation}
R_1+R_2+R_3=0
\end{equation}
and anomaly would not appear as it enables us to set the right hand
sides of equations in Eq.(11) to zero together. So, in QED$_{1+3}$,
 {\em the appearance of anomaly is also due to the presence of a 
kind of unambiguous rational terms} in $\Gamma^5_{\rho\nu\mu}$ 
which come from $I_{\mu\nu}$, a logarithmically divergent integral 
as in QED$_{1+1}$ (see Eq.(2) and Eq.(3)). That is the reason for 
the appearance of anomaly which is independent of regularization 
procedure and hence definite. 

To further support our viewpoint, we would like to list the 
results of the massive case with photons on shell ($p^2=q^2=0$)
\begin{eqnarray}
\Gamma^5_{\rho\nu\mu}
&=&\displaystyle\frac{e^2}{2\pi^2}
   \left \{ \epsilon_{\alpha\beta\nu\mu}
   q^{\alpha}p^{\beta}   \left [
   \displaystyle\frac{2q_{\rho}-p_{\rho}}{2p\cdot q}
   -\displaystyle\frac{\ln^2 (\sigma+1)/(\sigma-1)}{8\tau^2 p\cdot q}
   p_{\rho}-\displaystyle\frac{\sigma\ln (\sigma+1)/(\sigma-1)}
   {2p\cdot q} q_{\rho} \right ] \right. \nonumber\\
& &\ \ \ \ \ +\epsilon_{\alpha\beta\rho\mu}p^{\alpha}q^{\beta} \left [
   \displaystyle\frac{2p_{\nu}-q_{\nu}}{2p\cdot q}
   -\displaystyle\frac{\ln^2 (\sigma+1)/(\sigma-1)}{8\tau^2p\cdot q}
   q_{\nu}-\displaystyle\frac{\sigma\ln (\sigma+1)/(\sigma-1)}
   {2p\cdot q} p_{\nu} \right ] \nonumber\\
& &\ \ \ \ \ +\left .\epsilon_{\alpha\rho\nu\mu}(p-q)^{\alpha} \left [
    -\displaystyle\frac{\ln^2 (\sigma+1)/(\sigma-1)}{8\tau^2}+
     C_1-2C_2-C_3 \right ] \right \}, \nonumber\\
\Gamma^5_{\rho\nu}
&=&\displaystyle\frac{e^2}{4\pi^2}\epsilon_{\alpha\beta\rho\nu}
   p^{\alpha}q^{\beta}\displaystyle\frac{\ln^2 (\sigma+1)/(\sigma-1)}
    {4m^2\tau^2}
\end{eqnarray}
with
\begin{equation}
\tau=\sqrt{p\cdot q/(2m^2)},\ \ \ \sigma=\sqrt{1-1/\tau^2},
\end{equation}
and $C_1$, $C_2$ and $C_3$ being the three unknown constants.
Subsequently, we see
\begin{eqnarray}
q^{\rho}\Gamma^5_{\rho\nu\mu}
&=&\displaystyle\frac{e^2}{4\pi^2}\epsilon_{\alpha\beta\nu\mu}
   p^{\alpha}q^{\beta}C_{[\rho]},\nonumber\\
p^{\nu}\Gamma^5_{\rho\nu\mu}
&=&-\displaystyle\frac{e^2}{4\pi^2}\epsilon_{\alpha\beta\rho\mu}
   p^{\alpha}q^{\beta}C_{[\nu]},\nonumber\\
k^{\mu}\Gamma^5_{\rho\nu\mu}
&=&-2m \Gamma^5_{\rho\nu}-\displaystyle\frac{e^2}{4\pi^2}
   \epsilon_{\alpha\beta\rho\nu}p^{\alpha}q^{\beta}C_{[\mu]},
\end{eqnarray}
with
\begin{equation}
C_{[\rho]}=C_{[\nu]}=1+2(C_1-2C_2-C_3),\ \ \
C_{[\mu]}=4(2C_2+C_3-C_1).
\end{equation}
From (17) we have
\begin{equation}
C_{[\rho]}+C_{[\nu]}+C_{[\mu]}=2
\end{equation}
which is much the same of eq.(12). Again it is derived from the
existence of an unambiguous rational term in the ambiguous 
$I_{\mu\nu}$.

Now we write the chiral amplitude (in QED$_{1+1}$ or QED$_{1+3}$) 
in the following form (with the above experience)
\begin{equation}
\Gamma^5=\Gamma^5_{irra}+\Gamma^5_{ra}+\Gamma^5_{poly},
\end{equation}
where $\Gamma^5_{irra}$ refers to the irrational and well-defined 
part which obeys canonical relations, $\Gamma^5_{ra}$ represents 
the unambiguous rational part and $\Gamma^5_{poly}$ denotes the 
ambiguous polynomial of momenta with unknown coefficients. We 
should note that such a classification is based on our treatment 
offered in the beginning of this letter, {\em it is a hindsight 
rather than a new regularization}.
$\Gamma^5_{irra}$ contributes to $2m\Gamma^5_{\mu\nu}$ 
(see Eq.(16)),
while $\Gamma^5_{ra}$ and $\Gamma^5_{poly}$ contribute to the
$C_{[\mu]}$'s and to eq.(18).
{\em $\Gamma^5_{poly}$ can only drive anomaly around the vector 
and axial vector vertices but never remove it}. As shown above in 
Eq.(13), without the rational term, anomaly would disappear in 
suitable regularizations. Thus, it is the rational part which is 
the quantum mechanical origin of anomaly, i.e., it violating the 
canonical Ward identities. We feel that the rational part should 
be a definite measurement or reflection of some
physics hidden in the so-called short-distance singularity. (The 
main goal of a regularization in this problem should be to 
calculate out the rational part as well as the polynomial part 
from the potentially divergent Feynman integrals rather than 
merely separating the divergence out.) To our knowledge, this fact 
has never been explicitly demonstrated in QED$_{1+1}$ and 
QED$_{1+3}$ before. Conventionally, axial anomaly is
attributed to shift effects of internal momenta in relevant 
Feynman integrals that are linearly divergent. However, a little 
work will show that these effects just reside in the ambiguous 
polynomial part employing our new approach. According to the above 
discussions, there is little hope to find a "good" regularization 
to get rid of anomaly.

Now some remarks are in order. (a) As is well known, the 
Wess-Zumino term \cite{WZ}, which is a local functional of gauge 
fields and auxiliary scalar fields, can accommodate the anomaly. 
Here we see that anomaly comes from a nonlocal functional 
($\Gamma^5_{rat}$ is nonlocal) of gauge fields only. Thus it 
naturally leads us to take this nonlocal functional as another 
representation of Wess-Zumino action without resorting to scalar 
fields. In the simpler $1+1$-dimensional case, one may integrate 
out the scalar fields to get the nonlocal functional from
the local one \cite{JR}. Similar thing should be tractable in
$1+3$-dimensional case. (b) The famous Adler-Bardeen theorem 
\cite{ABA} about chiral anomaly, from our point of view, can now 
be understood as that there is no more similar rational terms in 
higher order radiative corrections. On the other hand, one may 
manipulate a proof of the Adler-Bardeen theorem along this line. 
The $1+1$-dimensional case is much simpler, where $e$ is 
dimensional, so higher order corrections can not yield a term like 
$e^n p_{\mu}p_{\nu}/p^2$ which is prohibited by dimensional 
analysis. We expect the rationality of the rational term
might be helpful in further understanding the structure of QFT. 
(c) In 1+2-dimensional case, there is no suitable structure of 
rational term that may lead to current anomaly. The Chern-Simons 
term is known to come from a decoupling limit of a definite term 
\cite {YY}. (d) It is worthwhile to note that anomaly must have 
come from a certain kind of rational term, but that does not mean 
all rational terms originate anomalies, the structure of rational 
term is important. The relation between the trace anomaly and the 
rational term is also established \cite {YN,Ta}. 

Recently, a reinvestigation of this problem in configuration space 
\cite {CSC} whose conclusion confirms ours here. The stress of 
conformal symmetry in the space-time expression of relevant 
amplitudes just corresponds to our stress of the unambiguous 
rational term which is a momentum space expressing of the 
conformal behavior as we originally noted \cite {YN}. And the 
regularization independence of the sum of the coefficients at the 
vector and axial vector vertices is also appreciated in \cite {CSC}. 
We want to note that our approach make us able to easily identify 
the unambiguous conformal term even if one does the calculation 
in massive case.
 
In conclusion, we have performed a more detailed and explicit 
study on the chiral anomaly problem within perturbative QFT 
framework. It is shown that the definite or quantum mechanical 
(regularization independent) source for the appearance of anomaly 
is closely related to the presence of a kind of rational term in 
momentum space (coming from a kind of logarithmically divergent 
Feynman integral). We expect the same situation remains valid in 
higher dimensional QFT. 

\vspace{1.0cm}
This work was supported jointly by the National Science Foundation
in China and the Science Foundation of State Education Commission 
in China.



\begin{thebibliography}{99}
\bibitem{Ja} see, e.g. R. Jackiw, in: \it Current Algebra and 
        Anomalies, \rm eds. S.B Treiman, R. Jackiw, B. Zumino and 
         E. Witten, (Princeton University Press, Princeton, 
         N.J.,1985), p.211 .
\bibitem{ABJ} S.L. Adler, Phys. Rev. 177 (1969) 2426;
              J.S. Bell and R. Jackiw, Nuovo Cimento 60 A (1969) 47;
              W.A. Bardeen, Phys. Rev. 184 (1969) 1848. 
\bibitem{DF}  M. D\"{u}tsch, F. Krahe and G. Scharf,
              Phys. Lett. B 258 (1991) 457;
              D.Z. Freedmann, K. Johnson and J.I. Lattore,
              Nucl. Phys. B 371 (1992) 353.
\bibitem{YN}  J.-F. Yang, Ph D Thesis, Fudan University, 1994, 
              unpublished;
              J.-F. Yang and G.-J. Ni, Acta Physica Sinica 4 (1995)
              88.
\bibitem{JF}  J.-F Yang, hep-th/9708104.
\bibitem{WZ}  J. Wess and B. Zumino, Phys. Lett. B37 (1971) 95. 
\bibitem{JR}  R. Jackiw and R. Rajaramann, Phys. Rev. Lett. 54 (1985)
              385;
              J. Paris, Phys. Lett. B 300 (1993) 104. 
\bibitem{ABA} S.L. Adler and W.A. Bardeen, Phys. Rev. 182
              (1969) 1517. 
\bibitem{YY}  J.-F. Yang and G.-J. Ni, Phys. Lett. B343 (1995) 249.
\bibitem{Ta}  G.-J. Ni and J.-F. Yang, Phys. Lett. B393 (1997) 79.
\bibitem{CSC} J. E. Erlich and D. Z. Freedman, MIT-CTP-2588, 
              hep-th/9611133, (1996) and references therein.
\end{thebibliography}
\end{document}